\documentclass[english,onecolumn,draftclsnofoot,12pt]{IEEEtran}
\usepackage[T1]{fontenc}
\usepackage[latin9]{inputenc}
\usepackage{amstext}
\usepackage{amssymb}
\usepackage{graphicx}

\makeatletter

\newcommand{\lyxmathsym}[1]{\ifmmode\begingroup\def\b@ld{bold}
  \text{\ifx\math@version\b@ld\bfseries\fi#1}\endgroup\else#1\fi}

\providecommand{\tabularnewline}{\\}

\usepackage{cite}

\usepackage{algorithm}

\usepackage{algpseudocode}

\makeatother

\usepackage{babel}
\begin{document}

\title{Mobile Converged Networks: Framework, Optimization and Challenges}

\author{%
\begin{tabular}{c}
Tao Han$^{\dagger}$,~\IEEEmembership{Member,~IEEE,} Yang Yang$^{\ddagger}$,~\IEEEmembership{Senior Member,~IEEE,}\tabularnewline
Xiaohu Ge$^{\dagger}$,~\IEEEmembership{Senior Member,~IEEE,} Guoqiang
Mao$^{\lyxmathsym{§}*}$,~\IEEEmembership{Senior Member,~IEEE}\tabularnewline
\tabularnewline
{\small{}$^{\dagger}$Department of Electronics and Information Engineering,
Huazhong University of Science and Technology, China}\tabularnewline
{\small{}$^{\ddagger}$School of Information Science and Technology,
ShanghaiTech University, China}\tabularnewline
{\small{}$^{\lyxmathsym{§}}$School of Computing and Communications,
University of Technology Sydney, Australia}\tabularnewline
{\small{}$^{*}$National ICT Australia (NICTA), Australia}\tabularnewline
{\small{}Email: \{hantao, xhge\}@hust.edu.cn, yangyang@shanghaitech.edu.cn,
g.mao@ieee.org}\tabularnewline
\end{tabular}\thanks{Accepted by IEEE Wireless Communications Magazine SI on Mobile Converged
Networks.}\thanks{Copyright (c) 2014 IEEE. Personal use of this material is permitted.
However, permission to use this material for any other purposes must
be obtained from the IEEE by sending a request to pubs-permissions@ieee.org.}\thanks{Digital Object Identifier 10.1109/MWC.2014.7000969}}
\maketitle
\begin{abstract}
In this paper, a new framework of mobile converged networks is proposed
for flexible resource optimization over multi-tier wireless heterogeneous
networks. Design principles and advantages of this new framework of
mobile converged networks are discussed. Moreover, mobile converged
network models based on interference coordination and energy efficiency
are presented and the corresponding optimization algorithms are developed.
Furthermore, future challenges of mobile converged networks are identified
to promote the study in modeling and performance analysis of mobile
converged networks.
\end{abstract}

\section{Introduction}

\begin{figure}[bh]
\begin{centering}
\includegraphics[width=0.8\columnwidth]{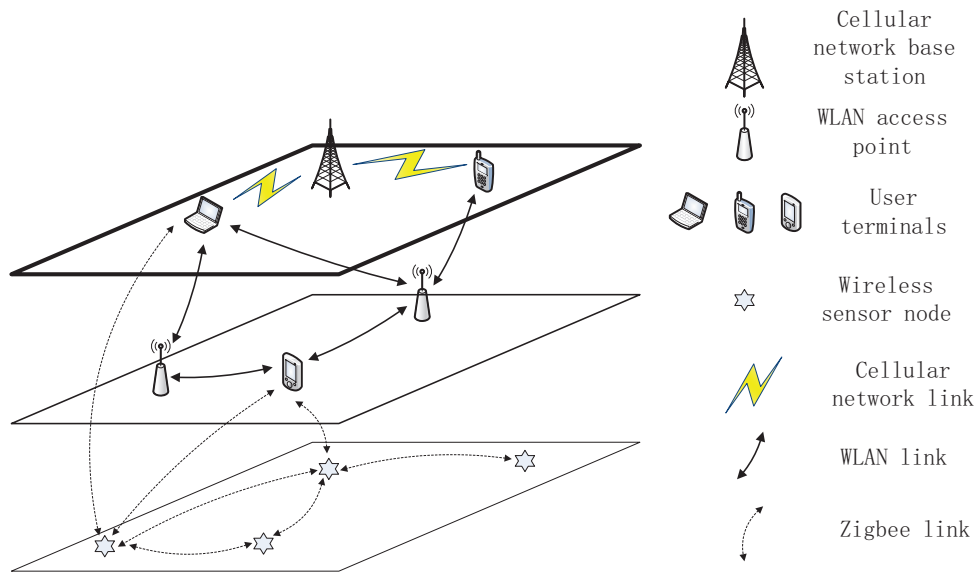}
\par\end{centering}

\caption{A typical mobile converged network\label{fig:A-typical-MCN}}
\end{figure}

Currently, wireless communication networks are widely deployed into
every scenario of human society, which includes short distance transmission,
e.g., Bluetooth communications in wireless mouses, long distance transmission,
e.g., space transmission between the Earth and the Mars, and etc.
Although the above networks and application scenarios are quite different,
multi-mode communications have adopted more and more communication
standards and technologies into a common user terminal. For instance,
the smart mobile phone usually supports the cellular network, the
wireless local area network (WLAN), the Bluetooth and the near field
communication, etc. Based on multi-mode communications, different
types of wireless heterogeneous networks can be converged into a uniform
mobile network, i.e., the mobile converged network\cite{Meddour2009,zhang2012}.
A typical mobile converged network is illustrated by Fig. \ref{fig:A-typical-MCN},
in which the cellular network, WLAN and wireless sensor network (WSN)
are converged.

Studying mobile converged networks has attracted much attention in
the past years, especially in the topic of converged mobile cellular
networks and WSNs. Han proposed an authentication and key agreement
protocol that efficiently reduces the overall computational and communication
costs in the next generation converged network\cite{han2011}. A system
architecture and application requirement for converged mobile cellular
networks and WSNs was introduced and then the joint optimization of
converged networks for machine-to-machine communications was discussed
in\cite{Shan2012}. An energy-efficient data collection scheme in
a converged WSN and mobile cellular network was proposed for improving
energy efficiency\cite{xiao2012}. A quality of service (QoS)-guaranteed
resource scheduling algorithm and a railway resource grab mechanism
in high-speed environment was presented to ensure the QoS for users
and the timely transmission for railway signal in mobile converged
networks\cite{gao2013}. Bae proposed a new concept of the converged
service based on the digital multimedia broadcast and wireless networks\cite{bae2013}.
After considering the network operation cost, the performance tradeoff
between the network quality of service and the network operation cost
for the intersystem soft handover in the converged digital video broadcasting
for handhelds and Universal Mobile Telecommunications System network
was modeled using a stochastic tree\cite{yang2008}. A framework of
combining clouds and distributed mobile networks was presented \cite{taleb2013mnet},
which the mobile converged networks can take advantage of. An efficient
network selection mechanism was presented to guarantee mobile users
selecting a most appropriate wireless network to connect from heterogeneous
wireless networks using the theory of games\cite{han2012}.

In all the aforementioned mobile converged networks studies, modeling
and performance analysis for detail scenarios were discussed and most
of mobile converged networks were consisted of two types of wireless
networks. However, there is not a general framework of mobile converged
networks which covers multi-tier wireless heterogeneous networks.
Motivated by above gaps, we propose a new framework of mobile converged
networks which not only covers main characteristics of mobile converged
networks but also includes different types of wireless heterogeneous
networks.

The remainder of this paper is outlined as follows. A new framework
of mobile converged networks is proposed to support multi-tier heterogeneous
networks. Moreover, two algorithms are developed to improve the performance
of mobile converged networks, respectively. Based on presented results,
future challenges for mobile converged networks are given, followed
by conclusions drawn in the last section.

\section{A Framework of Mobile Converged Networks}

\subsection{A framework of mobile converged networks}

To converge different types of wireless communication technologies
into a mobile network, two problems have to be solved. Firstly, a
converged scheme should be presented by investigating characteristics
of different types of wireless communication technologies for avoiding
potential technology conflicts. Secondly, the performance evaluation
of mobile converged networks is another issue which involves with
the selection of evaluation subjects, the evaluation approaches, the
validation of evaluation, the analysis of evaluation data, etc.

Based on the aggregation of each tier model of heterogeneous network,
the mobile converged network framework can be directly built by defining
the node multi-associated relationship in multi-tier heterogeneous
networks. However, it is difficult to analyze and optimize the resource
schedule of mobile converged networks when many variables are involved
into the aggregation of each tier model of heterogeneous networks.
Moreover, it becomes impossible to optimize the resource schedule
in mobile converged networks when the calculation complexity increases
obviously with the increasing of network sizes. Therefore, it is a
critical problem to build a new framework of mobile converged networks
with the limited complexity and enough flexibility for future wireless
heterogeneous networks.

In order to reduce the number of variable types in converged heterogeneous
networks, the variable mapping is considered as an important approach
to build a framework of mobile converged networks. For example, transmitters
in different tiers of wireless heterogeneous networks usually have
different transmission power. To evaluate the impact of transmission
power from different tier wireless heterogeneous networks on the mobile
converged networks, locations of transmitters can be scaled into a
framework of mobile converged network by accounting for path loss
effect over wireless channels. In the new framework of mobile converged
networks, the transmission power from difference tiers of wireless
heterogeneous networks is normalized and locations of transmitters
are scaled to ensure that the signal-to-interference-plus-noise ratio
(SINR) at receivers in a mobile converged network are equivalent to
the SINR at receivers in multi-tier wireless heterogeneous networks.
As a consequence, a new framework of mobile converged networks can
be presented by variable mapping in wireless heterogeneous networks.
Moreover, based on the allocation, the interaction of different tiers
of wireless heterogeneous networks is classified as the non-conflict
domain and the conflict domain in the framework of mobile converged
networks for optimizing the resource schedule. A conflict domain includes
multi-tier heterogeneous networks with interference among each other,
and a non-conflict domain includes multi-tier heterogeneous networks
without interference among each other.

\begin{figure}
\begin{centering}
\includegraphics[width=0.8\columnwidth]{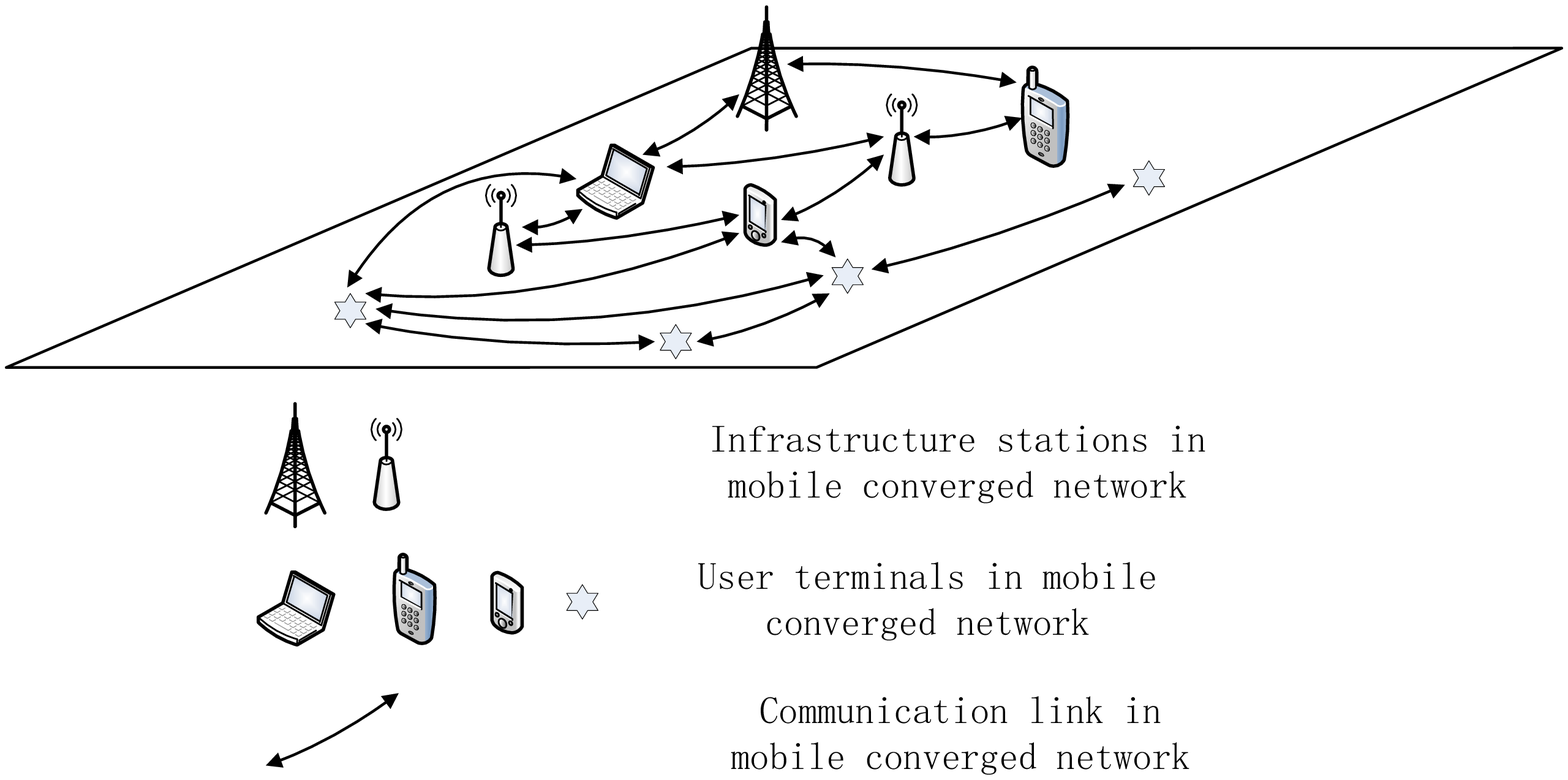}
\par\end{centering}

\caption{A random network topology of mobile converged network\label{fig:A-uniform-network-topology-of-MCN}}
\end{figure}

The network topology of wireless heterogeneous networks is affected
by wireless access methods, infrastructures, mobility and relay of
user terminals. Therefore, all above factors should be included into
the new framework of mobile converged networks by mapping a uniform
network topology. Based on the random network topology mapping approach,
a uniform network topology of mobile converged network is illustrated
in Fig. \ref{fig:A-uniform-network-topology-of-MCN}. Even for nodes
which have the multi-mode capability, different links of a multi-mode
node are mapped into virtual multi-tier wireless heterogeneous networks
as multiple links which are separated into multiple mapped single-mode
nodes with different transmission locations firstly. And then, using
virtual infinite-bandwidth inter-node links to connect them together,
these single-mode nodes at virtual multi-tier wireless heterogeneous
networks are mapped as a functional single node into a uniform network
topology of mobile converged network.

In the uniform network topology of mobile converged network, all infrastructure
stations and user terminals are normalized as mobile converged network
nodes and all wireless links of wireless heterogeneous networks are
normalized as mobile converged network links. Therefore, the uniform
network topology of mobile converged networks is formulated as $\left(\Phi^{C},L^{C}\right)$,
where $\Phi^{C}$ is the node set of mobile converged networks and
$L^{C}$ is the link set of mobile converged networks. Compared with
aggregated multi-tier wireless heterogeneous networks, the performance
analysis of mobile converged networks is simplified by a uniform network
topology.

For example, we consider the down-links in $K$-tier heterogeneous
cellular networks, the base stations (BSs) in tier $k$ have the transmit
power $P_{k}$ and the distribution of BSs follows a point process
\cite{stoyan1995stochastic} $\Phi_{k}^{\text{BS}}$. Without loss
of generality, a mobile station (MS) $\text{MS}_{y}$ is assumed to
be located at the origin coordinate $\mathbb{O}$, the signal power
$P_{\text{BS}_{x}\text{MS}_{y}}$ received by the MS $\text{MS}_{y}$
from a BS $\text{BS}_{x}$ located at $x$ is 
\begin{equation}
P_{\text{BS}_{x}\text{MS}_{y}}=P_{k}\left\Vert x-\mathbf{\mathbb{O}}\right\Vert ^{-\alpha}=1\cdot\left(P_{k}^{-\frac{1}{\alpha}}\left\Vert x-\mathbf{\mathbb{O}}\right\Vert \right)^{-\alpha}=1\cdot\left\Vert P_{k}^{-\frac{1}{\alpha}}\cdot x-\mathbb{O}\right\Vert ^{-\alpha},\label{eq:power_tpnm}
\end{equation}
 where $1$ is the normalized transmission power, $\alpha$ is the
path loss factor from the BS $\text{BS}_{x}$ to the MS $\text{MS}_{y}$
at origin $\mathbf{\mathbb{O}}$ and $\left\Vert \cdot\right\Vert $
is the distance operator.

The equation (\ref{eq:power_tpnm}) means that the signal power received
by the MS $\text{MS}_{y}$ at origin is exactly equal to the signal
power transmitted from a virtual BS with transmit power $1$ and located
at $P_{k}^{-\frac{1}{\alpha}}\cdot x$, where $P_{k}^{-\frac{1}{\alpha}}$
times far from the origin $\mathbb{O}$ comparing to the BS $\text{BS}_{x}$.
Centered with the origin $\mathbb{O}$, which we call transmission
power normalized center, the point process $\Phi_{k}^{\text{BS}}$
can be scaled to the point process $\Phi_{k}^{\text{BS}\prime}=P_{k}^{-\frac{1}{\alpha}}\cdot\Phi_{k}^{\text{BS}}$,
in which the virtual BSs with transmission power $1$ transmit the
same signal power (or interference power) to the origin MS as the
original BSs in $\Phi_{k}^{\text{BS}}$ do. We scale the point processes
in all tiers to normalize the transmit power to $1$, and then combine
them into a point process $\Phi^{C}=\bigcup_{k=1}^{K}P_{k}^{-\frac{1}{\alpha}}\cdot\Phi_{k}^{\text{BS}}$,
while keeping the link set $L^{C}$ same as the network links before
scaling. In the end, the uniform network topology of mobile converged
networks is given.

\begin{figure}
\begin{centering}
\includegraphics[width=0.8\columnwidth]{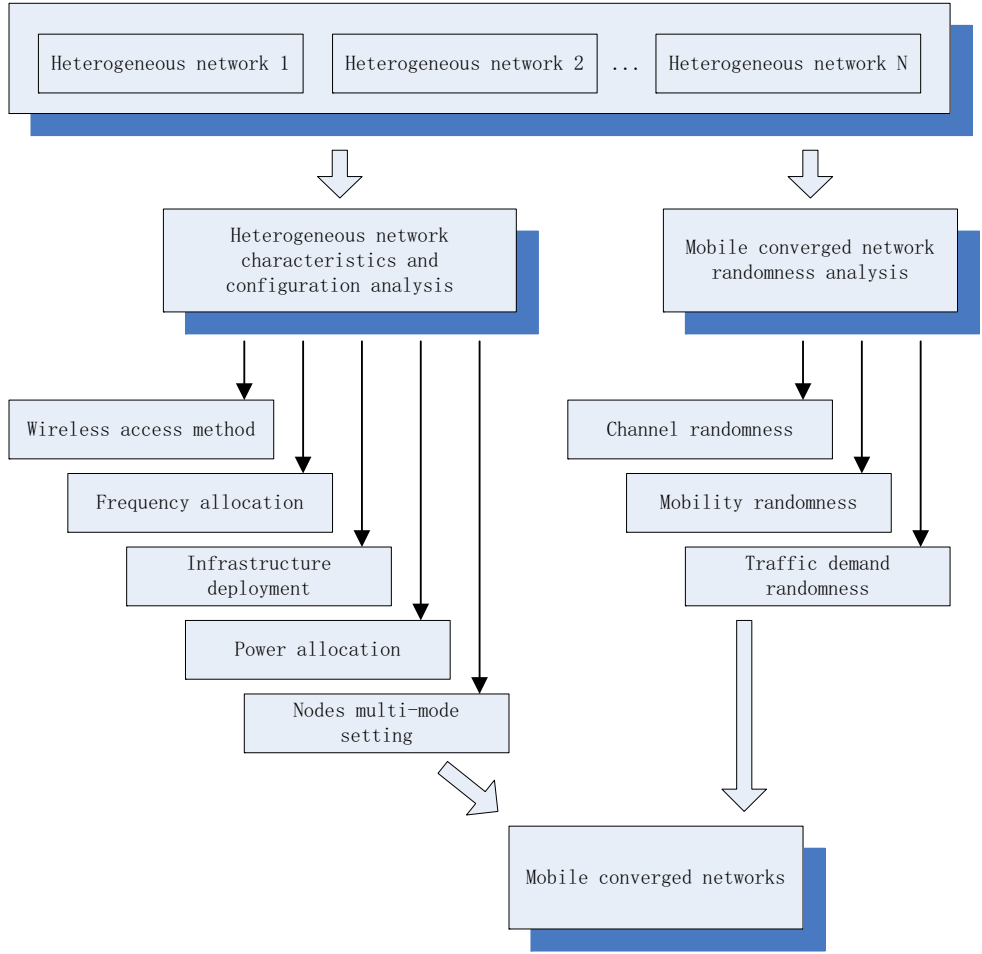}
\par\end{centering}

\caption{The new framework of mobile converged networks\label{fig:The-new-framework-of-MCNs}}
\end{figure}

However, there still exist many issues to build the new framework
of mobile converged networks. Two main issues are summed as follows: 
\begin{enumerate}
\item The uniform description issue of different wireless access methods.
It is well known that the wireless access method has great impact
on the wireless link management, the interference coordination, the
transmission rate, spectrum efficiency and energy efficiency in wireless
networks. It is difficult to analyze the impact of different wireless
access methods on the performance of mobile converged networks by
one parameter. 
\item The location issue of transmission power normalized center. In the
view of statistics, the transmission power normalized center of mobile
converged networks can located in arbitrary point without impacting
of signal/interference analysis on the mobile converged network. However,
different locations of the transmission power normalized center will
affect the random network topology and network router models. Consequently,
location optimization of the transmission power normalized center
is another key issue in the framework of mobile converged networks.
\end{enumerate}
Motivated by the above gaps, we propose a new framework of mobile
converged networks to cover multi-tier wireless heterogeneous networks
in Fig. \ref{fig:The-new-framework-of-MCNs}. The new framework of
mobile converged networks is divided into two parts: one part of framework
is about the characteristics and configurations of wireless heterogeneous
networks, which includes wireless access methods, frequency allocation,
infrastructure deployment, power allocation and configuration of multi-mode
nodes; another part of framework is about the randomness characteristics
of wireless heterogeneous networks, which includes wireless channel
randomness, mobility randomness and traffic demands randomness. 

\subsection{Modeling of mobile converged networks based on interference coordination}

The interference coordination is always an important challenge for
wireless communication systems. It is well known that the interference
coordination is related with the capacity, transmission rate, spectrum
efficiency and energy efficiency in wireless networks. Based on the
proposed framework of mobile converged networks, we explore to build
a mobile converged network model accounting for co-channel interference
among multi-tier wireless heterogeneous networks.

\begin{figure}
\begin{centering}
\includegraphics[width=0.55\columnwidth]{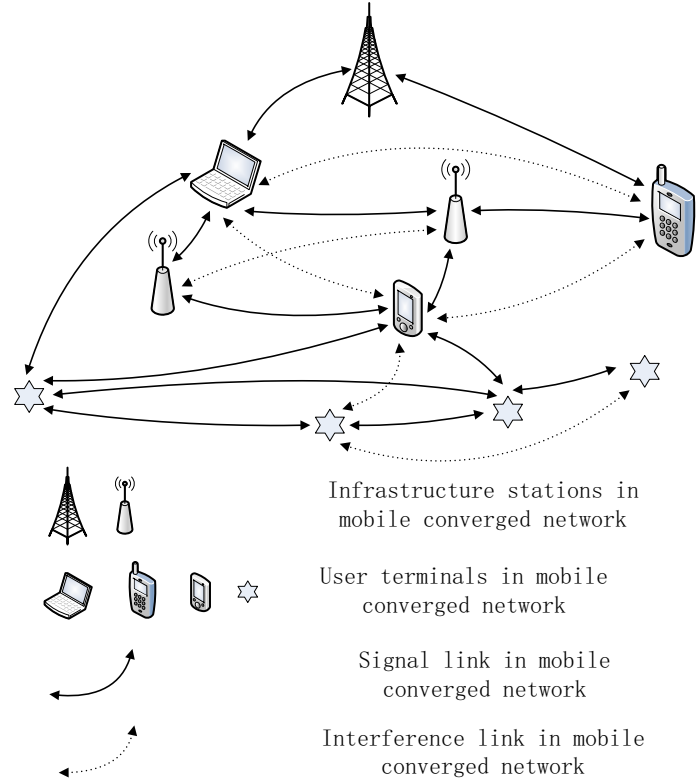}
\par\end{centering}

\caption{Network topology of mobile converged networks based on interference
coordination\label{fig:Network-topology-of-MCNs-b_o-interf-coord}}
\end{figure}

Based on the proposed framework of mobile converged networks, multi-tier
wireless heterogeneous networks are mapped into a mobile converged
network with a uniform network topology. However, the uniform network
topology is just based on links in wireless heterogeneous networks.
The links affected by co-channel interference of mobile converged
networks are not mapped into the uniform network topology. For the
mobile converged network model based on interference coordination,
node links of mobile converged networks are adjusted by relationships
between interfering transmitters and receivers. Furthermore, the coordinate
locations of the transmitters in mobile converged networks are scaled
by considering the transmission power of interfering transmitters
and distances between the interfering transmitters and receivers.
The random topology of mobile converged networks based on interference
coordination is illustrated in Fig. \ref{fig:Network-topology-of-MCNs-b_o-interf-coord},
where real lines denote desired signal links and dash lines denote
interference links.

Considering the interference coordination, the uniform network topology
with desired signal links and interference links is derived for mobile
converged networks. Furthermore, the cross-tier routing algorithms
can be developed by utilizing the relay capability of multi-mode nodes
for minimizing co-channel interference in mobile converged networks.
The main idea of cross-tier routing algorithms is to maximize distances
between interfering transmitters and receivers. To satisfy this requirement,
data traffic is relayed by multi-mode nodes with different frequency
bands in a mobile converged network. In this case, the co-channel
interference could be minimized with guaranteed throughput in mobile
converged networks. Therefore, the joint optimization solution involves
with frequency, spatial and temporal dimensionalities of mobile converged
networks. However, the mobile converged network model based on interference
coordination is an open question considering following challenges:
1) Time variant wireless channels. Time variant wireless channels
affect the capacity and the bit error rate of wireless links and even
interrupt wireless links in mobile converged networks. In this case,
the topology of mobile converged networks is affected by time variant
wireless channels. Accordingly, the interference model of mobile converged
networks should support the dynamical network topology affected by
time variant wireless channels. 2) Spatial randomness of interfering
transmitters. In mobile converged networks, the associating relationship
among multi-mode nodes is flexible for minimizing co-channel interference.
However, it is a complex problem to build the interference conflicting
model when the network topology of mobile converged networks is changed
by relationships between multi-mode nodes and user terminals. 3) Mobility
randomness of user terminals. In mobile converged networks, user terminal
locations are usually assumed to follow a random process. Moreover,
the mobility model of user terminals also follows another random process.
However, it is very difficult to model the mobility of the user terminals,
because the mobility model and the location model follow different
random processes.

Based on the mobile converged network model with interference coordination,
the interference minimizing algorithm of mobile converged networks
is presented in Algorithm \ref{algo:interf-minimizing}.

\begin{algorithm}
\caption{the interference minimizing algorithm of mobile converged networks}  
\label{algo:interf-minimizing}
\begin{algorithmic}[1]
\State Input initial configuration parameters in mobile converged networks, such as locations of nodes, transmission power and resource allocation;
\State Conflicting evaluation in frequency, spatial and temporal domains of mobile converged networks;
\While {True}
  \State Evaluate conflicting distribution in multi-domains;
  \State Calculate the aggregated interference $I_C$;
  \Repeat
    \State $I_{OPT} \gets I_C$
    \State Search the region with the maximal co-channel interference;
    \State Check the interference tolerance in every tier of wireless heterogeneous network, and then adjust the cross-tier routing into the tier of network with the maximal interference tolerance;
    \State Evaluate the new conflicting distribution in frequency, spatial and temporal domains;
    \State Calculate the new aggregated interference $I_C$;
  \Until {$I_C \ge I_{OPT}$}
  \State Revise the configuration parameters based on nodes mobility;
\EndWhile
\end{algorithmic}
\end{algorithm}

\begin{figure}
\begin{centering}
\includegraphics[width=0.9\columnwidth]{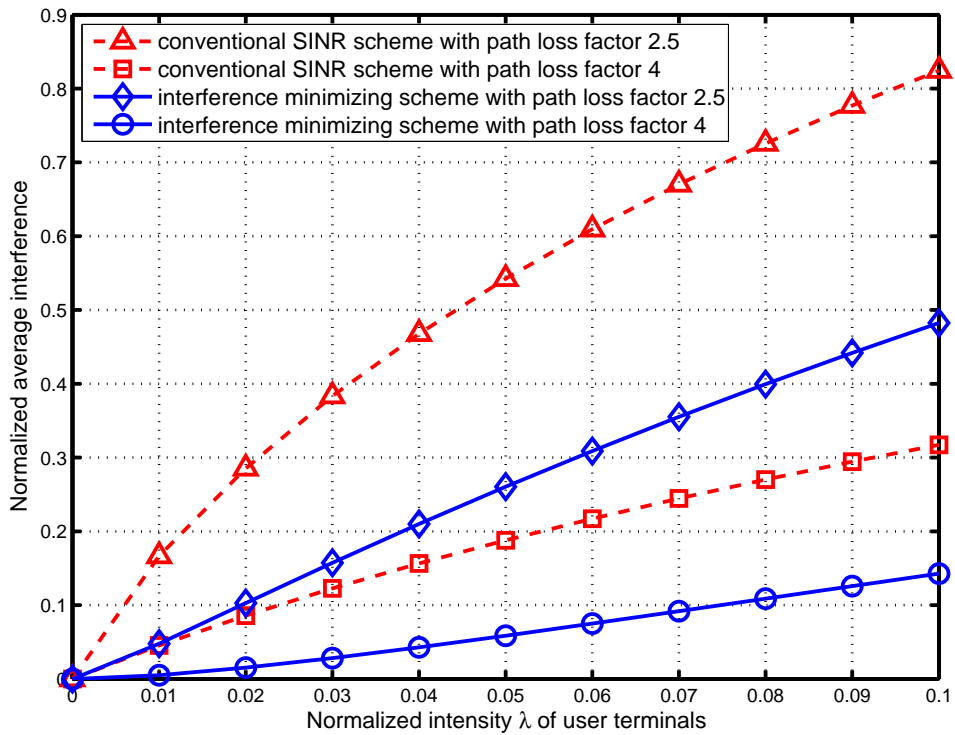}
\par\end{centering}

\caption{Normalized average interference of mobile converged networks with
interference minimizing scheme and conventional SINR scheme\label{fig:Interf-min-in-3-tier-CN}}
\end{figure}

In conventional SINR scheme of mobile converged networks, user terminals
are associated with infrastructure stations based on the maximal SINR
value received at the user terminal. In this case, every user terminal
just associates with an infrastructure station based on itself SINR
performance without considering interference at other user terminals.
As a consequence, one of user terminals maybe obtains the best optimal
performance, but the average interference of mobile converged networks
is obviously increased. Contrary to the conventional SINR scheme,
the Algorithm \ref{algo:interf-minimizing} try to decrease the average
interference at user terminals of mobile converged networks to improve
the total performance of mobile converged networks. In every loop
of interference optimization, the cross-tier route for transmission
detours to the tier in which the smallest interference is generated
and the maximal interference tolerance remains. Based on the Algorithm
\ref{algo:interf-minimizing}, simulation results of the interference
minimizing scheme and conventional SINR scheme are compared in Fig.
\ref{fig:Interf-min-in-3-tier-CN}. In simulations of Fig. \ref{fig:Interf-min-in-3-tier-CN},
a three-tier wireless heterogeneous network is configured with corresponding
infrastructure station density denoted as 1:0.1:0.01, and the intensity
of user terminals is also normalized by these densities of the tiers.
User terminals are governed by a Poisson point process with intensity
$\lambda$ and their mobility subjects to Gaussian-Markov mobility
model \cite{liang2003tnet}, and every user terminal can access one
tier of wireless heterogeneous networks based on different association
schemes. From curves in Fig. \ref{fig:Interf-min-in-3-tier-CN}, the
normalized average interference of mobile converged networks with
the interference minimizing scheme is less than the normalized average
interference of mobile converged networks with the conventional SINR
scheme under different path loss factors. Fig. \ref{fig:Interf-min-in-3-tier-CN}
illustrates that the normalized average interference increases with
the increasing of the normalized intensity of user terminals, because
the shorter distance of user terminals conduces to the higher interference.

\subsection{Modeling of mobile converged networks based on energy efficiency}

Energy efficiency of wireless networks is defined as how many bits
can be transmitted by consuming one Joule energy in wireless networks.
Because mobile converged networks have characteristics of multi-network
spatial overlay and multi-mode configuration in transmitters, mobile
converged networks are recommended as one of potential solutions for
improving energy efficiency in future wireless networks. Heterogeneous
characteristics of mobile converged networks provide many degrees
of freedom to optimize the energy efficiency of data transmission
\cite{ge2014tvt}.

When multi-tier wireless heterogeneous networks are mapped into a
uniform network topology, weighting vectors are added into links of
the uniform network topology to build a directed graph. On the other
hand, we can also configure the maximal power consumption threshold
in specified nodes of mobile converged networks. It is necessary to
build the maximal network flow model considering multi-dimensionality
randomness in mobile converged networks. Furthermore, the energy efficiency
optimization solution of mobile converged networks is based on the
trade-off of the energy consumption and the network flow in frequency,
spatial and temporal dimensionalities \cite{xiang2013twc}. The Algorithm
\ref{algo:e-e-optimization} is the energy efficiency optimization
algorithm of mobile converged networks, in which energy consumption
of every possible cross-tier link is evaluated to maximize the energy
efficiency of mobile converged networks.

\begin{algorithm}
\caption{the energy efficiency optimization algorithm of mobile converged networks}  
\label{algo:e-e-optimization}
\begin{algorithmic}[1]
\State Input initial configuration parameters in mobile converged networks, such as locations of nodes, transmission power and resource allocation;
\State Calculate the energy efficiency and the network flow in frequency, spatial and temporal dimensionalities of mobile converged networks;
\While {True}
  \State Calculate the network flow $F_C$ and the energy efficiency $E_C$;
  \Repeat
    \State $E_{OPT} \gets E_C$
    \State Search the minimal energy efficiency region in a mobile converged network;
  \State Analyze the network flow in multi-tier heterogeneous networks and corresponding energy efficiency;
    \State Adjust the network flow by cross-tier routing algorithms to maximize the energy efficiency;
    \State Calculate the new energy efficiency $E_C$ in the specified region;
  \Until {$E_C \ge E_{OPT}$}
  \State Change the network topology based on the mobility of user terminals;
\EndWhile
\end{algorithmic}
\end{algorithm}

\begin{figure}
\begin{centering}
\includegraphics[width=0.9\columnwidth]{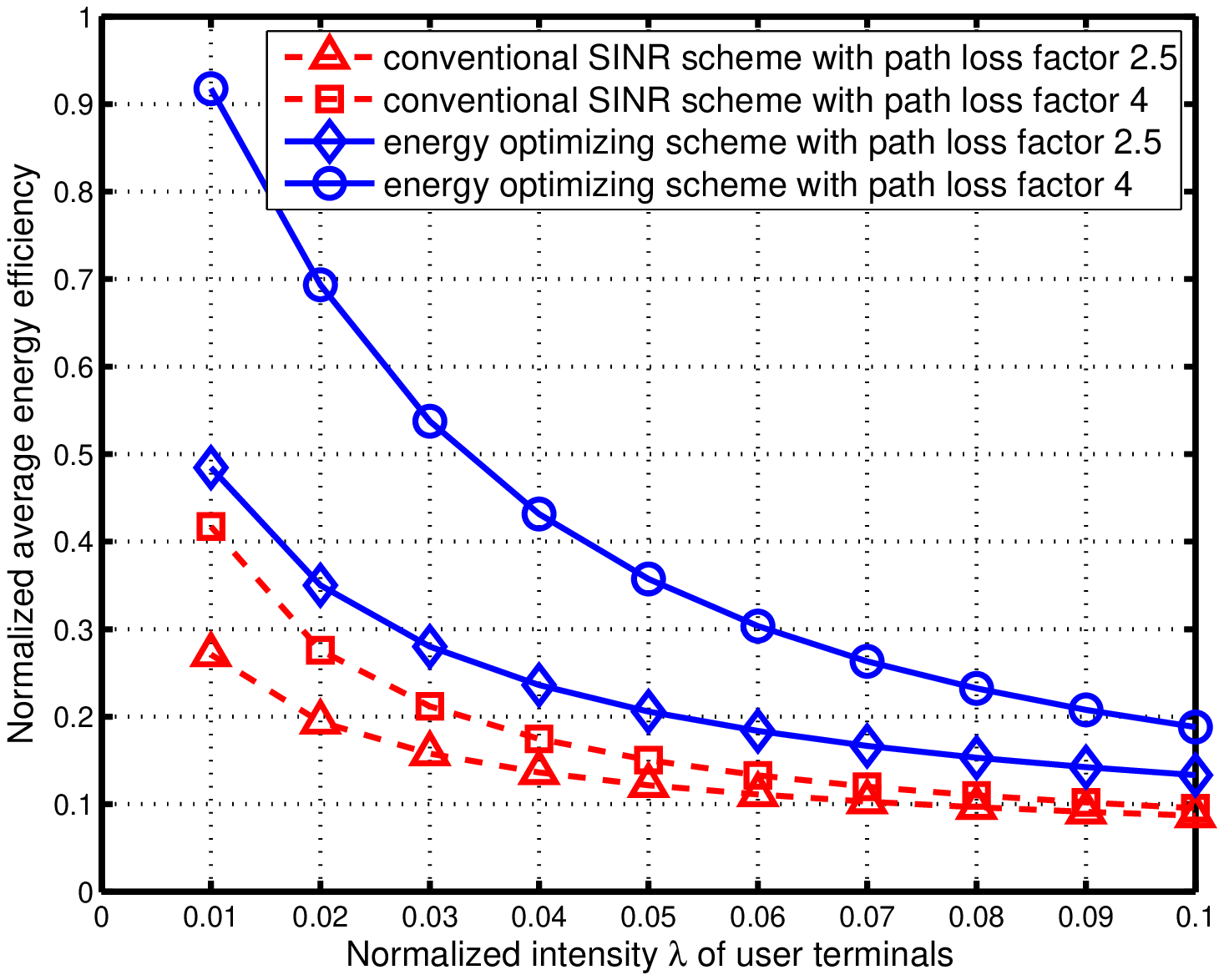}
\par\end{centering}

\caption{Normalized average energy efficiency of mobile converged networks
with energy efficient scheme and conventional SINR scheme\label{fig:Opt_EE_in_MCN}}
\end{figure}

To evaluate the performance of the proposed energy efficiency optimization
algorithm, simulation results are compared between the conventional
SINR scheme and the energy efficiency optimization scheme for mobile
converged networks in Fig. 6. The system model and simulation parameters
in Fig. \ref{fig:Opt_EE_in_MCN} are configured as the same in Fig.
\ref{fig:Interf-min-in-3-tier-CN}. Based on curves in Fig. \ref{fig:Opt_EE_in_MCN},
the normalized energy efficiency of mobile converged networks with
the energy efficiency optimization scheme is larger than the normalized
energy efficiency of mobile converged networks with the conventional
SINR scheme under different path loss factors. Fig. \ref{fig:Opt_EE_in_MCN}
shows that the normalized average energy efficiency decreases with
increasing of the normalized intensity of user terminals, because
the higher interference conduces to the lower capacity and the lower
energy efficiency.

The interference coordination model and the energy efficiency model
are originated from the proposed framework of mobile converged networks.
Based on the different constraint metrics, the interference coordination
model can be used for interference optimization and the energy efficiency
model can be used for energy optimization in mobile converged networks.
Moreover, based on other resource optimization constraints in the
framework of mobile converged networks, many models can be built for
performance analysis.

\section{Future Challenges}

It is always a great challenge to build a comprehensive system model
of mobile converged networks to cover different types of characteristics
in multi-tier wireless heterogeneous networks. Such a characteristic
usually cannot be represented by a single system parameter. In addition,
accuracy and effectiveness are equally important for a comprehensive
system model, otherwise the computational complexity of such a model
will increase dramatically and cannot be tackled by typical performance
evaluation approaches. So, what are the key trade-offs between accuracy,
effectiveness and complexity in mobile converged network models is
an interesting question for future research. For a particular system
model, there remains many open research issues, such as design and
deployment of multi-tier network architecture, node position optimization,
and resource optimization across multiple networks, in the analysis
of system performance under realistic network conditions.

Based on the proposed framework and the comprehensive system model,
new algorithms can be developed to improve the performance of mobile
converged networks. Besides interference coordination and energy efficiency,
different applications and user behaviors, such as audio/video streaming,
interactive games, and online news, and the corresponding QoS provisioning
and the resource allocation will also affect the whole system performance
of mobile converged networks. Therefore, new resource optimization
algorithms considering throughput, delay and link reliability should
be further developed and analyzed accounting for different application
types and scenarios.

Although there exist some obstacles for operators and researchers
to evaluate performance of mobile converged networks considering the
different standards and commerce security, this is not an excuse not
to converge multi-tier wireless heterogeneous networks into a mobile
converged network. To ensure this outcome, the standards of mobile
converged networks measurement and estimation should be made a matter
of regulation and enforcement by the regulatory authorities.

\section{Conclusion}

Until recently, the convergence of different types of heterogeneous
networks becomes one of main research directions in future wireless
networks. In this paper, we propose a new framework of mobile converged
networks to cover different types of heterogeneous networks. A uniform
framework of mobile converged networks would be helpful to design
and evaluate performance of mobile converged networks in a single
model. Furthermore, considering objectives of interference coordination
and energy efficiency, two models of mobile converged networks and
corresponding algorithms are developed. However, there still exist
many issues to converge different types of heterogeneous networks,
such as modeling, resource optimization and performance evaluation
of mobile converged networks. If these are done, a veritable challenge
would indeed emerge in the next round of the telecommunications revolution.

\section*{Acknowledgments}

The corresponding author of the article is Prof. Xiaohu Ge. The authors
would like to acknowledge the support from the International Science
and Technology Cooperation Program of China (Grant No. 2014DFA11640,
2012DFG12250 and 0903), the National Natural Science Foundation of
China (NSFC) (Grant No. 61471180, 61271224 and 61301128), the NSFC
Major International Joint Research Project (Grant No. 61210002), the
Hubei Provincial Science and Technology Department (Grant No. 2013BHE005),
the Fundamental Research Funds for the Central Universities (Grant
No. 2011QN020, 2013ZZGH009 and 2014QN155), and EU FP7-PEOPLE-IRSES
(Contract/Grant No. 247083, 318992 and 610524). Yang's work has been
partially supported by the NSFC under grant 61231009, the Ministry
of Science and Technology 863 program under grant 2014AA01A707, and
the National Science and Technology Major Projects under grant 2014ZX03005001.
This research is also supported by Australian Research Council Discovery
projects DP110100538 and DP120102030.

\bibliographystyle{ieeetr}
\bibliography{Paper_Ref_MWC_2014}

\begin{IEEEbiographynophoto}{Tao Han}
 {[}M'13{]} (hantao@hust.edu.cn) received the Ph.D. degree in Communication
and Information Engineering from Huazhong University of Science and
Technology (HUST), Wuhan, China in December, 2001. He is currently
an Associate Professor with the Department of Electronics and Information
Engineering, HUST. His research interests include wireless communications,
multimedia communications, and computer networks.
\end{IEEEbiographynophoto}

\begin{IEEEbiographynophoto}{Yang Yang}
 is currently a Professor with the School of Information Science
and Technology, ShanghaiTech University, and the Director of Shanghai
Research Center for Wireless Communications (WiCO). Prior to that,
he has served Shanghai Institute of Microsystem and Information Technology
(SIMIT), Chinese Academy of Sciences, as a Professor; the Department
of Electronic and Electrical Engineering at University College London
(UCL), United Kingdom, as a Senior Lecturer; the Department of Electronic
and Computer Engineering at Brunel University, United Kingdom, as
a Lecturer; and the Department of Information Engineering at The Chinese
University of Hong Kong as an Assistant Professor. His research interests
include wireless ad hoc and sensor networks, wireless mesh networks,
next generation mobile cellular systems, intelligent transport systems,
and wireless testbed development and practical experiments.
\end{IEEEbiographynophoto}

\begin{IEEEbiographynophoto}{Xiaohu Ge}
 {[}M'09-SM'11{]} (xhge@hust.edu.cn) is currently a Professor with
the Department of Electronics and Information Engineering at Huazhong
University of Science and Technology (HUST), China. He received his
Ph.D. degree in Communication and Information Engineering from HUST
in 2003. From January 2013, he was granted as a Huazhong scholarship
professor. He serves as an Associate Editor for the IEEE ACCESS, Wireless
Communications and Mobile Computing Journal, etc..
\end{IEEEbiographynophoto}

\begin{IEEEbiographynophoto}{Guoqiang Mao}
 {[}S'98-M'02-SM'08{]} received PhD in telecommunications engineering
in 2002 from Edith Cowan University. He currently holds the position
of Professor of Wireless Networking, Director of Center for Real-time
Information Networks at the University of Technology, Sydney. His
research interest includes intelligent transport systems, applied
graph theory and its applications in networking, wireless multihop
networks, wireless localization techniques and network performance
analysis.\end{IEEEbiographynophoto}

\end{document}